\documentclass[sigconf]{acmart}

\usepackage{comment}
\usepackage{caption}
\usepackage{subcaption}
\usepackage{multirow}   
\usepackage{multicol}   
\usepackage{url}

\graphicspath{ {./figs/} }

\AtBeginDocument{%
  \providecommand\BibTeX{{%
    \normalfont B\kern-0.5em{\scshape i\kern-0.25em b}\kern-0.8em\TeX}}}

\setcopyright{acmcopyright}
\copyrightyear{2018}
\acmYear{2018}
\acmDOI{10.1145/1122445.1122456}

\acmConference[SenSys '21]{SenSys '21: ACM Conference on Embedded Networked Sensor Systems}{November 03--05, 2021}{SenSys, Coimbra, Portugal}
\acmBooktitle{SenSys '21: ACM Conference on Embedded Networked Sensor Systems,
  November 03--05, 2021, SenSys, Coimbra, Portugal}
\acmPrice{15.00}
\acmISBN{978-1-4503-XXXX-X/18/06}

\begin{document}

\title{Dataset: Large-scale Urban IoT Activity Data for DDoS Attack Emulation}

\author{Arvin Hekmati}
\email{hekmati@usc.edu}
\affiliation{%
  \institution{University of Southern California}
  \city{Los Angeles}
  \state{California}
  \country{USA}
}

\author{Eugenio Grippo}
\email{egrippo@usc.edu}
\affiliation{%
  \institution{University of Southern California}
  \city{Los Angeles}
  \state{California}
  \country{USA}
}

\author{Bhaskar Krishnamachari}
\email{bkrishna@usc.edu}
\affiliation{%
  \institution{University of Southern California}
  \city{Los Angeles}
  \state{California}
  \country{USA}
}

\renewcommand{\shortauthors}{A. Hekmati, E. Grippo, B. Krishnamachari}

\begin{abstract}

    \textbf{  As IoT deployments grow in scale for applications such as smart cities, they face increasing cyber-security threats. In particular, as evidenced by the famous Mirai incident and other ongoing threats, large-scale IoT device networks are particularly susceptible to being hijacked and used as botnets to launch distributed denial of service (DDoS) attacks. Real large-scale datasets are needed to train and evaluate the use of machine learning algorithms such as deep neural networks to detect and defend against such DDoS attacks. We present a dataset from an urban IoT deployment of 4060 nodes describing their spatio-temporal activity under benign conditions. We also provide a synthetic DDoS attack generator that injects attack activity into the dataset based on tunable parameters such as number of nodes attacked and duration of attack. We discuss some of the features of the dataset. We also demonstrate the utility of the dataset as well as our synthetic DDoS attack generator by using them for the training and evaluation of a simple multi-label feed-forward neural network that aims to identify which nodes are under attack and when. }

\end{abstract}



\maketitle

\section{Introduction}

    Technological evolution has made possible the deployment of large internet of thing (IoT) systems that are able to connect multiple different sensors and actuators, allowing them to communicate and exchange enormous amounts of data. Such large-scale IoT systems consisting of thousands of sensor nodes are being proposed, for example, in the context of smart-city applications (\cite{7555867}, \cite{8276827}, \cite{nassar2017survey}). As these IoT systems grow in size and complexity, they are increasingly vulnerable from a Cybersecurity perspective (\cite{lu2018internet}, \cite{alferidah2020iot}, \cite{alrashdi2019iot}). 
    
    Most significantly, they are liable to be hacked into and hijacked by malicious entities and then used as part of massive botnets as a launching ground for distributed denial of service (DDoS) attacks, potentially affecting millions of end-users (\cite{bertino2017botnets}, \cite{kolias2017ddos}, \cite{hallman2017ioddos}, \cite{10.1016/j.jnca.2017.02.009}, \cite{garcia2014survey}, \cite{zargar2013survey}).
    A famous example of an IoT-based DDoS attack was the Mirai botnet, first identified in August  2016 by MalwareMustDie, a whitehat security research group. Afterward, some of the biggest DDOS attacks in history were performed by Mirai botnet and its mutated variants. 400,000 nodes infected by this malware executed DDoS attacks on websites with a massive peak of 1.1 Tbps data transfer (\cite{antonakakis2017understanding}, \cite{sinanovic2017analysis}, \cite{vengatesan2018analysis}, \cite{kolias2017mirai}, \cite{mcdermott2018mirai}, \cite{ margolis2017miari}, \cite{8538636}).

    One of the major steps to defend against such DDoS attacks is to detect them successfully, as close to real-time as possible. To this end, security researchers have turned to machine learning tools such as deep learning networks for DDoS attack detection (\cite{sharma2016machine}).
    The latest development in technology (parallel computing, high computational processing speeds, GPU’s, etc.) have made possible the vertiginous development of deep neural networks (NN's) (\cite{gurney2018introduction}, \cite{alpaydin2014introduction}, \cite{zayegh2018neural}), however their performance is very much dependent on the availability of rich datasets to train them.  The training of NN models will be challenged in the near future as botnet attacks become increasingly complex and grow in volume infecting massive portions of IoT networks. Therefore, it is important to build and maintain suitable large-scale IoT dataset repositories, useful for training such models, to allow the research community to stay ahead of malicious attack developments.

    There have been some prior efforts on collecting or creating synthetic DDoS attack datasets that could be used for training data-driven deep learning models (e.g., \cite{UNB}, \cite{UCI}, \cite{UNSW}, \cite{9099844}, \cite{ring2017flow}, \cite{ring2017creation_two}). However many of these studies either are not specifically focused on IoT devices (e.g., \cite{CAIDA}, \cite{sharafaldin2019developing}, \cite{DARPA}, \cite{UNB_two}, \cite{UNB_three}, \cite{UNB_four}, \cite{mci/Beer2017}, \cite{erhan2020bougazicci}), while others present data from a limited number of nodes (e.g., \cite{bot_iot}, \cite{UCI_two}, \cite{UNSW_two}). Further, while today's DDoS attacks such as Mirai are often relatively easy to detect because their traffic volumes dramatically exceed normal traffic (e.g., \cite{meidan2018mirai} reports nearly 100\% true positives for an autoencoder-based detection scheme on the Mirai botnet), we believe that future attacks will be more cleverly camouflaged so that the traffic generated during the attack from a given node matches the traffic volume from a benign node. Thus, beyond datasets that show the behavior of today's attacks, there is still a need for large-scale datasets that characterize the benign activity of real IoT devices, which could be used as a basis to emulate more challenging future attacks. 
    
    
    We present in this work first a data set obtained from a real urban IoT system in a large city consisting of more than 4000 spatially distributed sensors. The data consists of the binary activity status of each node at a granularity of 30 seconds over a period of one month under a benign (non-attacked) setting. 
    
    To make the dataset useful for training machine learning tools for DDoS detection, we need to augment it through synthetic attack emulation.  Therefore, in addition to the raw (benign) activity data, we also provide a dedicated script that generates attacks in the proposed dataset synthetically. Our script allows the setting of multiple parameters: number of nodes to be processed, total attack duration, attack ratio, starting time of the attack; and each particular node is equipped with a time-stamp and an output that varies binary between 0 (no attack) and 1 (attacked node). To illustrate the utility of our dataset and attack emulator, we design, train, and implement a simple supervised feed-forward NN model to detect malicious attacks utilizing the provided dataset. We make the dataset and the attack emulation script along with our illustrative NN model available as an open-source repository online at \url{https://github.com/ANRGUSC/Urban_IoT_DDoS_Data}.

    The paper is structured as follows: section \ref{sec:original} presents the original dataset and some statistics about it. The attack and defense mechanism are presented in section \ref{sec:attack}. Finally, we conclude the paper in section \ref{sec:conclusion}.

\section{Original and Benign Activity Datasets}
\label{sec:original}
    The original data has been collected from the activity status of real event-driven IoT nodes deployed in an urban area\footnote{The source of this data has been anonymized for privacy and security reasons.}. The original dataset contains three main features, the node ID, the location of the node in Latitude and Longitude, and a timestamp of the activity status of the IoT node. A record has been added to the original dataset whenever the activity status of a node changes. The raw dataset has 4060 nodes with one month worth of data.\\
    
    Having a record of each node whenever the status of that node changes provides a bias towards the information of nodes that have more activity changes during the day. In order to overcome this issue, we also provide a script  that takes the original dataset and generates a new benign activity dataset showing the activity status for each node every $t_s$ seconds. In this way, all nodes in the benign activity dataset will have the same number of records. The script can generate a customized benign dataset by providing the beginning and ending date, the number of IoT nodes, and the time step, i.e., $t_s$.

    \begin{figure}
        \centering
        \includegraphics[width=\columnwidth]{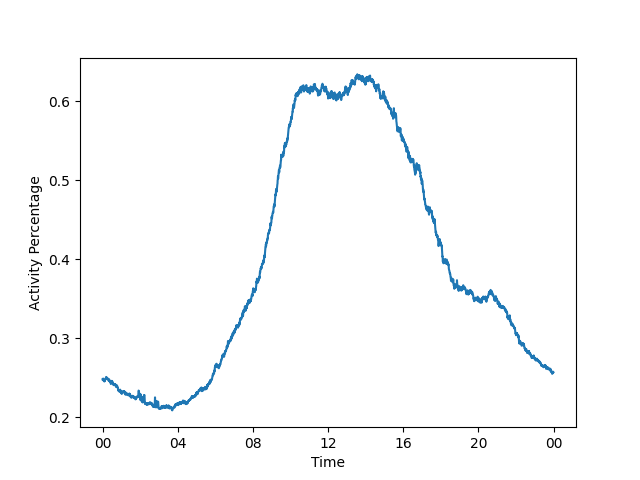}
        \caption{Active Nodes Percentage vs Time}
        \label{fig:active_nodes_percentage}
    \end{figure}

    \begin{figure}[]
      \centering 
      \includegraphics[width=\columnwidth]{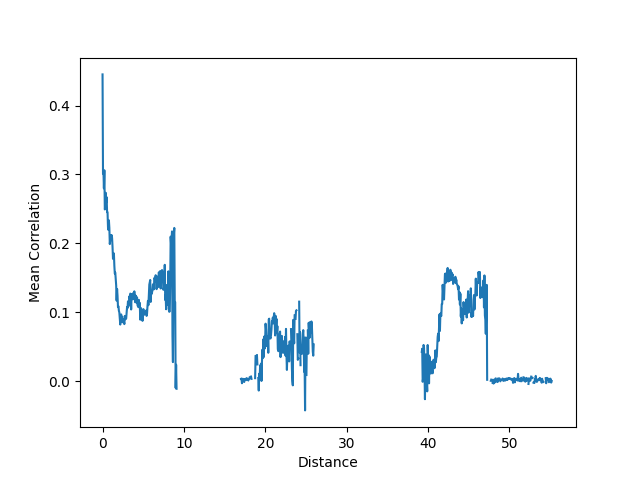}
      \caption{Nodes Activity Mean Correlation vs Distances}
      \label{fig:mean_pearson_correlation}
    \end{figure}

    \begin{figure*}[]
        \centering 
        \begin{subfigure}[]{0.45\textwidth}
          \includegraphics[width=\textwidth]{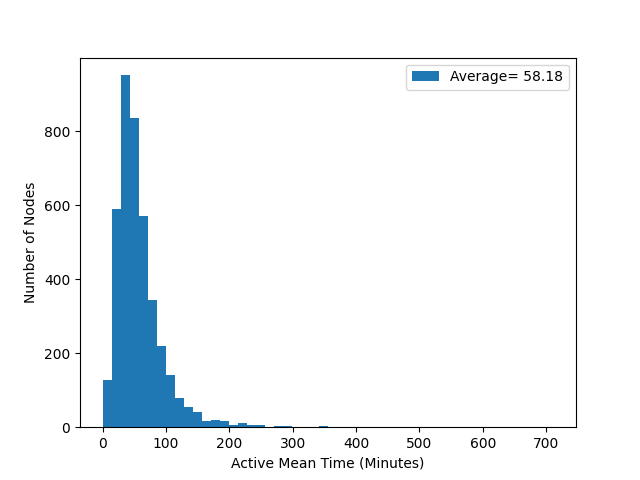}
          \caption{Histogram of mean activity time per node - From 8 AM to 8 PM}
          \label{fig:day_active_mean_time}
        \end{subfigure}\hfill
        \begin{subfigure}[]{0.45\textwidth}
          \includegraphics[width=\textwidth]{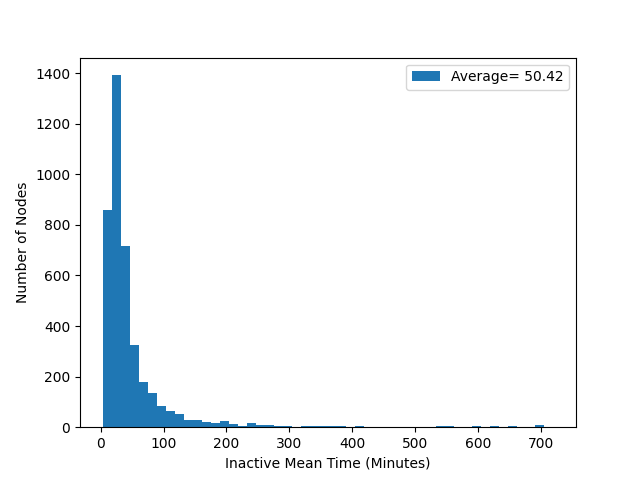}
          \caption{Histogram of mean inactivity time per node - From 8 AM to 8 PM}
          \label{fig:day_not_active_mean_time}
        \end{subfigure}\hfill
        \begin{subfigure}[]{0.45\textwidth}
          \includegraphics[width=\textwidth]{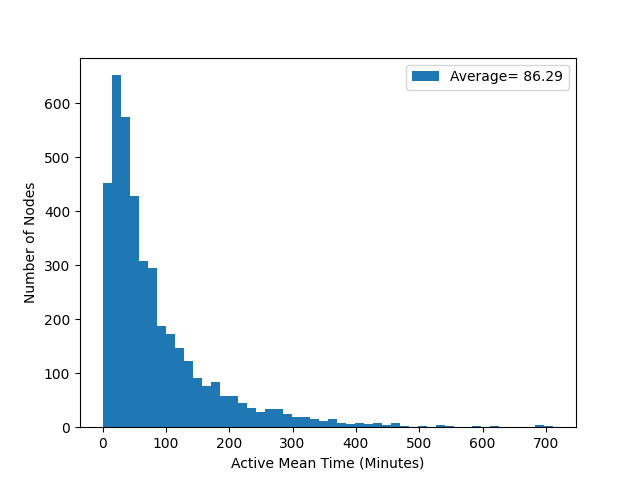}
          \caption{Histogram of mean activity time per node - From 8 PM to 8 AM}
          \label{fig:night_active_mean_time}
        \end{subfigure}\hfill
        \begin{subfigure}[]{0.45\textwidth}
          \includegraphics[width=\textwidth]{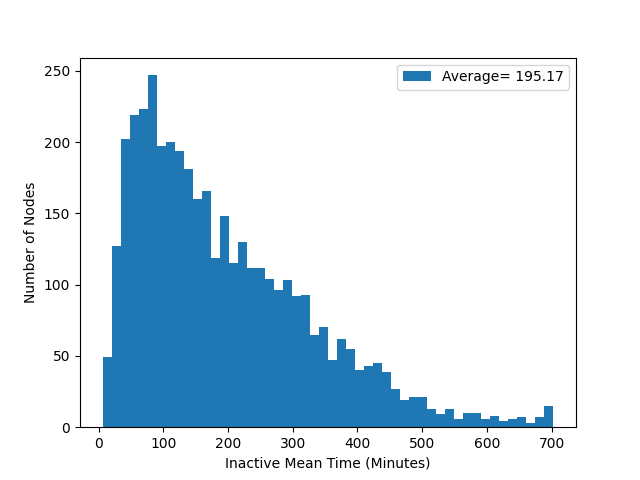}
          \caption{Histogram of mean inactivity time per node - From 8 PM to 8 AM}
          \label{fig:night_not_active_mean_time}
        \end{subfigure}
        \caption{Histograms of mean activity and inactivity time per node}
        \label{fig:active_mean_time}
    \end{figure*}
    
    \subsection{Dataset Statistics}
        In this subsection, some statistics of the datset are presented to illustrate the dataset's properties.
        
        Figure \ref{fig:active_nodes_percentage} presents the mean number of active nodes versus the time of the day on one particular day of the dataset. As we can see, up to 65\% of the nodes get activated around the middle of the day, but by midnight only about 20\% of the nodes are active.
        
        Figure \ref{fig:mean_pearson_correlation} shows the mean correlation between pairs of 500 randomly selected nodes in the dataset, in terms of their activity versus their distance. For this purpose, the pair distances between the nodes are calculated using Euclidean distance. The Pearson correlation has been used to calculate the correlation between nodes' activity in the one month worth of data. Then, the distances between nodes have been split into 1000 bins, and the average correlation for the nodes whose pair distances fall in the bin boundaries has been calculated for each bin. As we can see, generally, the farther the nodes are, their correlation will be lower (dropping from a mean correlation of about 0.4 for nodes that are very close to each other to values around 0.1 or below for distances greater than 10 units).
        
        Next, we present some statistics that show how long nodes tend to remain active or inactive. Figure \ref{fig:active_mean_time} shows a histogram of the mean active time for all nodes as well as the mean inactive time, for both the day time (8 AM to 8 PM) and nighttime (8 PM to 8 AM). Figures \ref{fig:day_active_mean_time} and \ref{fig:night_active_mean_time} show the active mean time for the day and night, respectively. Figures \ref{fig:day_not_active_mean_time} and \ref{fig:night_not_active_mean_time} show the inactive mean time for the day and night, respectively. As we can see in these figures, both mean active times and mean inactive times tend to be higher in the nights compared to the days.
    
\section{Attack and Defense Mechanism}
\label{sec:attack}
    
    This section presents how synthetic DDoS attacks are generated on the IoT nodes. Furthermore,  here we define the training dataset features and also the detection mechanism.

    \subsection{Generating Attack Dataset}
        In this paper, we synthetically generate a DDoS attack on the IoT nodes by setting all attacked nodes to an active status for the duration of the attack. Note that this approach to attack generation is more coarse-grained than providing the volume of packets or packets generated with specific content and destinations during an attack. We plan to incorporate the generation of such additional fine-grained information during the attacks in the near-future. But even by focusing on activity status only, we are effectively emulating more challenging futuristic DDoS attacks that may be hard to detect from a single node's traffic. 
        
        A script is provided that can be used for generating attacks over the dataset. Three parameters can be set in generating the attacks: the start time of the attack, the duration of the attack, and the percentage of the nodes that go under attack. In our experiments, we used one week's worth of dataset for generating attacks. The attacks are started at 2 AM on each day of the week over all of the nodes with durations of 1, 2, 4, 8, 16 hours.    

    \subsection{Generating Training Dataset}
        Given the attacked dataset, a labeled training dataset will be generated by calculating the mean activity time of the IoT nodes in the specified time windows. Note that, in the script provided for this paper, one could set the desired time windows for generating the training features. In this paper, we considered a list of 12 different time windows, namely as 1, 10, 30, minutes and 1, 2, 4, 8, 16, 24, 30, 36, 42, 48 hours, to calculate the mean activity time of the nodes. 
        
    \subsection{Defense Mechanism}
        
        We train a feed-forward neural network to detect the DDoS attack on the IoT nodes based on their collective activity status over time. As noted before, detecting attacks based on activity status alone is more challenging than approaches based on measuring fine-grained traffic volumes or flow-level information.  
        
        The model we have trained is a simple binary classification to show a sample usage of the presented dataset. In this neural network, we have an input layer with 12 neurons. The input layer is followed by one hidden layer with 8 neurons and ReLU activation. A dropout of 20\% and batch normalization is also used at the end of the hidden layer. The output is a single neuron with the Sigmoid activation function. In this experiment, we randomly selected 20 IoT devices nodes to generate attacks. We are training 20 different models for each IoT node using its data alone, each with different weights but all having the same architecture. The neural network model is trained for 500 epochs for each node to detect the attacked time slots in the dataset. We used one week's worth of data as the training dataset and another week as the testing dataset. Note that this is a simple approach that will not take into account any correlations in the data across different nodes, so there is scope for further improvement by developing more complex models that integrate the inputs from multiple IoT devices.

        \begin{table}[]
        \caption{Mean accuracy and recall for the NN detection model}
        \centering
        \begin{tabular}{|c|c|c|}
        \hline
        \textbf{}                 & \textbf{Mean Accuracy} & \textbf{Mean Recall} \\ \hline
        \textbf{Training Dataset} & 0.94                   & 0.93                 \\ \hline
        \textbf{Testing Dataset}  & 0.88                   & 0.84                 \\ \hline
        \end{tabular}
        \label{table:results}
        \end{table}  
        
\begin{figure}
            \centering
            \includegraphics[width=\columnwidth]{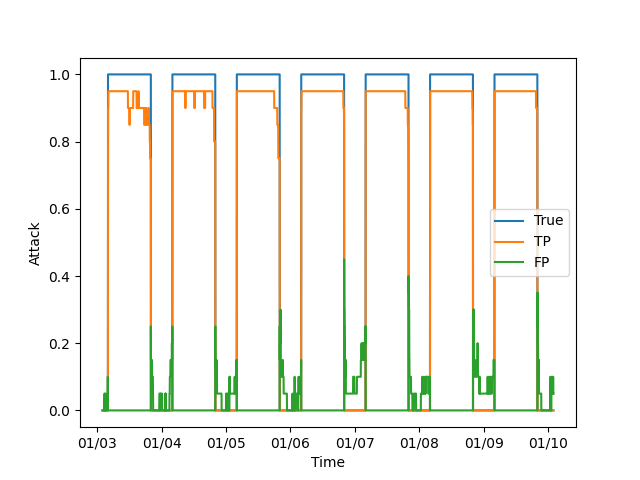}
            \caption{Training Dataset Attack Prediction vs Time}
            \label{fig:train_predict}
        \end{figure}

\begin{figure}
            \centering
            \includegraphics[width=\columnwidth]{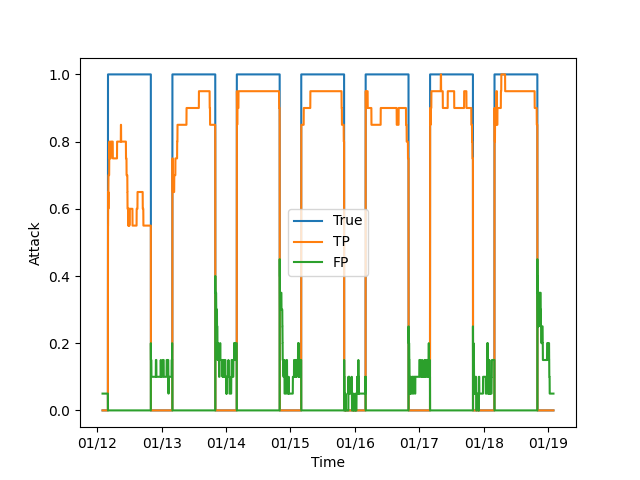}
            \caption{Testing Dataset Attack Prediction vs Time}
            \label{fig:test_predict}
        \end{figure}

       Table \ref{table:results} present the mean recall and accuracy of the 20 models trained for detecting DDoS attacks. 
       Figures \ref{fig:train_predict} and \ref{fig:test_predict} show the true attack attack (T), attack predictions true positive (TP) and false positives (FP) mean over all nodes vs time, for both training and testing dataset. In these figures we used the attack duration of 16 hours. As we can see, the attacked nodes are being detected very well in the training dataset with a few FP. On the other hand, we are getting around 84\% recall on the testing dataset with a little bit more FP.
        
\section{Conclusion}
\label{sec:conclusion}
  
    We have presented a new spatio-temporal dataset describing the activity of a 4060-node event-based urban IoT deployment. We have also provided a script to create a benign dataset out of the original dataset to reduce bias toward nodes with more activity. We have shown some statistical analyses on the dataset to illustrate some its key properties. We have also presented a synthetic DDoS attack generator to generate attacks using the dataset, and illustrate the training and evaluation of a feed-forward neural network using the dataset as a way to detect such attacks. We hope that the dataset and tools we have provided are helpful to the community to undertake various types of research related to large-scale event-driven IoTs, including DDoS attack detection.

\section{Acknowledgments}
    This material is based upon work supported by Defense Advanced Research Projects Agency (DARPA) under Contract No. HR001120C0160 for the Open, Programmable, Secure 5G (OPS-5G) program. Any views, opinions, and/or findings expressed are those of the author(s) and should not be interpreted as representing the official views or policies of the Department of Defense or the U.S. Government.


\bibliographystyle{unsrt} 
\bibliography{main} 


\end{document}